\documentclass[aps,twocolumn,showpacs,floatfix]{revtex4}

\usepackage{amsfonts}
\usepackage{amsmath}
\usepackage{amssymb}
\usepackage{graphicx}
\usepackage{color}
\usepackage{verbatim}
\usepackage{lipsum}
\usepackage{dsfont}
\usepackage{ulem}

\usepackage[T1]{fontenc}

\usepackage{hyperref}
\usepackage{makecell}
\usepackage{upgreek}

\usepackage{enumitem}

\begin{document}

\title{Tunable magnetic field source for magnetic field imaging microscopy}
\author{Andris Berzins$^1$}
\email{andris.berzins@lu.lv}
\author{Hugo Grube$^1$}
\author{Reinis Lazda$^1$}
\author{Marc A. Hannig$^{2,1}$}
\author{Janis Smits$^{3,1}$}
\author{Ilja Fescenko$^1$}

\affiliation{$^1$Laser Centre, University of Latvia, Jelgavas Street 3, LV-1004 Riga, Latvia}
\affiliation{$^2$Faculty of Mathematics and Physics, Leibniz University Hannover, 30167 Hannover, Germany}
\affiliation{$^3$The University of New Mexico, Albuquerque, United States}


\begin{abstract}
In this work we present a novel, compact, power efficient magnetic field source design for magnetic field imaging microscopy. The device is based on a pair of diametrically magnetized permanent magnet cylinders with electro-mechanical rotation control and ferrite field concentrators. A Hall probe and NV centers in diamond are used to demonstrate a proof of concept of a proposed magnetic field setup and to characterise the homogeneity of the produced magnetic field on a micrometer scale. Numerical simulation results are compared with experimental results showing good agreement of the distribution of the magnetic field in the setup. As a result, a magnetic field source with a tunable field amplitude in the range from 1~mT to 222~mT is demonstrated, achieving a magnetic field homogeneity of 2~ppm/$\upmu$m or 0.5~$\upmu$T/$\upmu$m at 222~mT in a 25$\times$25~$\upmu$m field of view.

\end{abstract}

\maketitle


\section{Introduction}
Nitrogen-Vacancy (NV) centers in diamond~\cite{ashfold_nitrogen_2020} have demonstrated themselves as a useful platform for magnetic field microscopy, that can be used to reveal two dimensional magnetic structures of various samples: magnetic mapping of superconducting thin films~\cite{waxman_diamond_2014, xu_mapping_2019} and microscopic electromagnetic devices~\cite{pham_magnetic_2011, mizuno_simultaneous_2020,horsley_microwave_2018}, as well as imaging of micrometer and nanometer sized particles~\cite{fescenko_diamond_2019,smits_estimating_2016}, thin film structures~\cite{lenz_imaging_2021,appel_nanomagnetism_2019,berzins_surface_2021,berzins_characterization_2021}, nanoscale magnetic resonance~\cite{ziem_quantitative_2019}, magnetic spins~\cite{steinert_magnetic_2013} and eddy currents~\cite{chatzidrosos_eddy-current_2019} to name a few. Furthermore, it is evident that NV based microscopy is well suited for the investigation of wide range temperature dependent magnetic effects on a microscopic scale as well~\cite{waxman_diamond_2014, wang_coherent_2020, plakhotnik_luminescence_2010}.

In many cases relatively strong magnetic fields are favored to reach the material's magnetic saturation in case of paramagnetic or superparamagnetic materials~\cite{fescenko_diamond_2019,berzins_surface_2021}, or to reach the anti-crossing regions of the NV centers~\cite{zhang_battery_2021}. However the devices that can reach the few hundred mT level often are electromagnets that are bulky, require either high voltage or high currents delivered with high accuracy, and are relatively expensive to make and exploit, as well as are prone to heating. With this research we present a solution for a magnetic field source for magnetic field imaging microscopy using NV centers in diamond that is based on rotatable permanent magnets in combination with magnetic field concentrators (ferrites). The presented setup can deliver magnetic fields up to 222~mT with a magnetic field homogeneity of down to 2~ppm/$\upmu$m at the 222~mT magnetic field configuration. Although this device is demonstrated using NV centres, it can be used for other applications as well. Here we provide a blueprint for a setup for a compact, variable and uniform magnetic field source, using cylindrical permanent magnets with the magnetization direction perpendicular to the cylindrical axis of the magnets.

Using such a setup allows for rapid changes from a close to zero magnetic field to a relatively high magnetic field by rotating both permanent magnets around their main cylindrical axes. Such a mode of operation gives a useful tool for hysteresis measurements, that usually require a combination of relatively large magnetic fields with the possibility to gradually increase or decrease the magnetic field, as well as change the field direction~\cite{sun_magnetic_2021,maertz_vector_2010,gu_flat-cladding_2018,paterson_measuring_2018}.
In order to increase the homogeneity of the magnetic field between the two permanent magnets, ferrite magnetic field concentrators are placed between the permanent magnets. The magnetic field concentrators improve the magnetic field homogeneity, and amend any small misalignment of the permanent magnets.

\section{Experimental methods}

To determine the uniformity of the magnetic field on micro-scale in the volume between the two permanent magnets we made Hall probe and NV based measurements of the spatial distribution of Optically Detected Magnetic Resonance (ODMR)~\cite{rondin_magnetometry_2014} on a diamond placed in between the magnet system.

\begin{figure}[h]
  \begin{center}
    \includegraphics[width=0.25\textwidth]{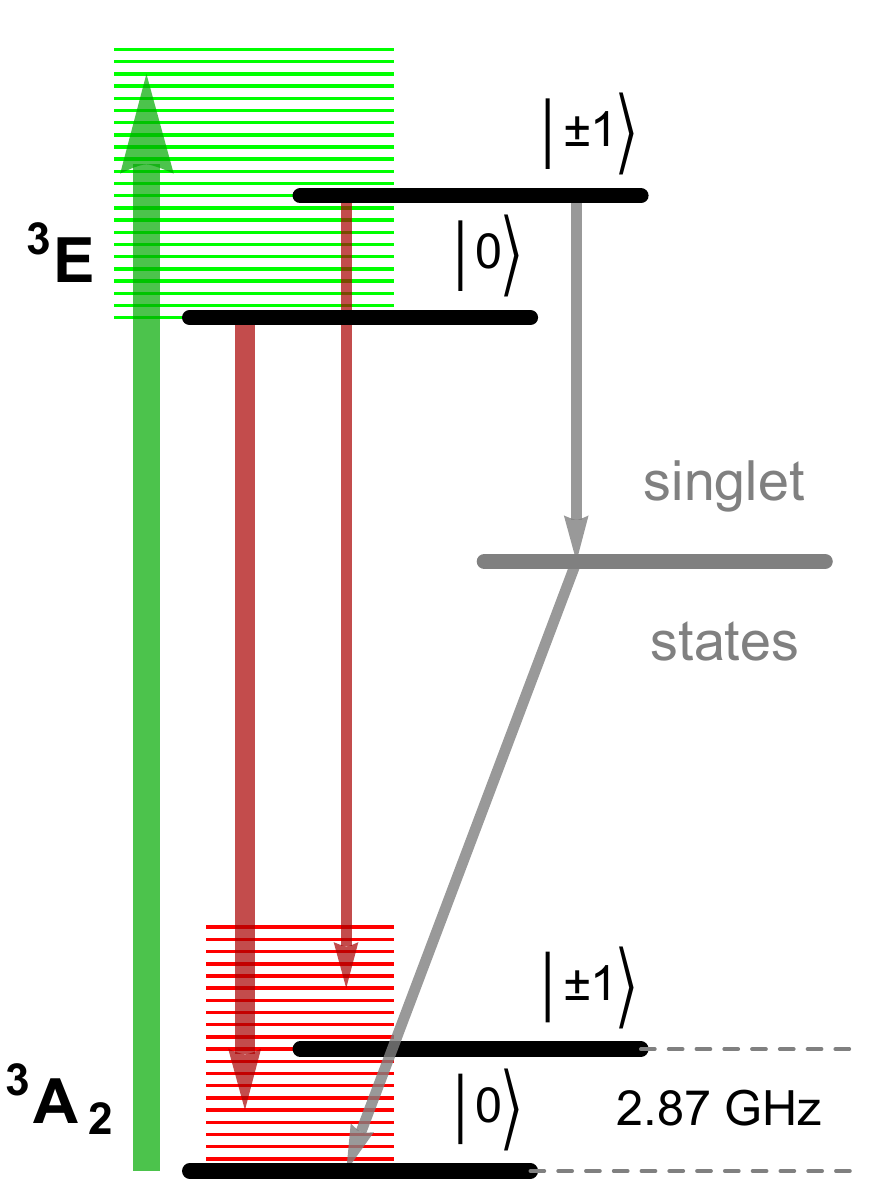}
  \end{center}
  \caption{NV energy level diagram depicting magnetic sublevels ($\vert0\rangle$, $\vert\pm1\rangle$), their phonon bands (green and red horizontal lines), optical excitation (green arrow), fluorescence (red arrows), and nonradiative (gray arrows) pathways. Gray arrows show spin-selective intersystem crossing leading to polarization into the \textbar0\big \rangle~ground-state sublevel.}
  \label{levels}
\end{figure}

The NV energy levels with optical excitation and emission pathways are depicted in Figure \ref{levels}. When green laser light (532 nm) continuously excites NV centers, the ground-state population is optically polarized into the $\vert0\rangle$ state by a non-radiative, spin-selective decay via intermediate singlet states.
Because of this spin-selective decay mechanism, NV centers excited from the $\vert0\rangle$ state emit fluorescence in the red region of the spectrum at a higher rate than those originating from $\vert\pm1\rangle$ states. Application of a microwave (MW) field mixes the spin populations, resulting in a decrease in the fluorescence when the MW frequency matches the spin transition frequencies, $f_{\pm}$. Evaluation of these frequencies by precisely measuring the NV fluorescence as the MW frequency is swept across the resonances is the basic principle of ODMR techniques.

As the luminescence signal was collected and detected using an optical system, that maintains the two-dimensional information of the luminescence distribution of the NV layer in the diamond, the magnetic field distribution over the field of view could be reconstructed after data processing (described in next section).

\section{Experimental setup}
\begin{figure*}
      \includegraphics[trim=20 5 5 5,clip,width=0.45\textwidth]{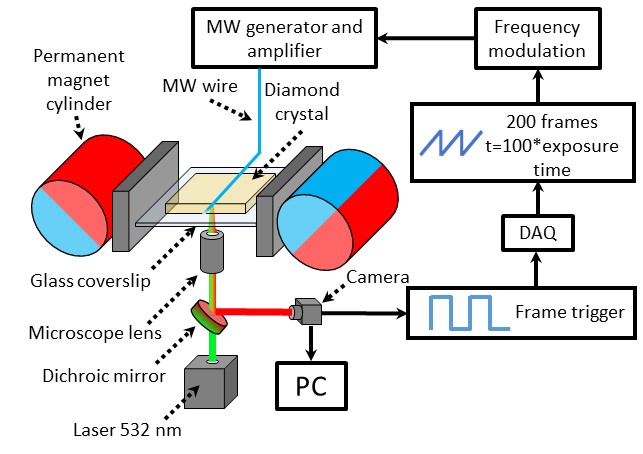}
      \includegraphics[width=0.30\textwidth]{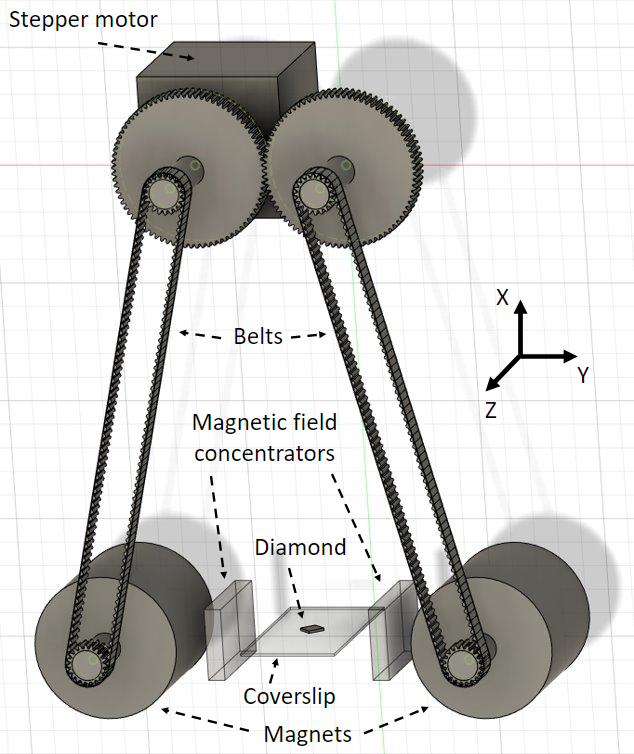}
      \includegraphics[width=0.23\textwidth]{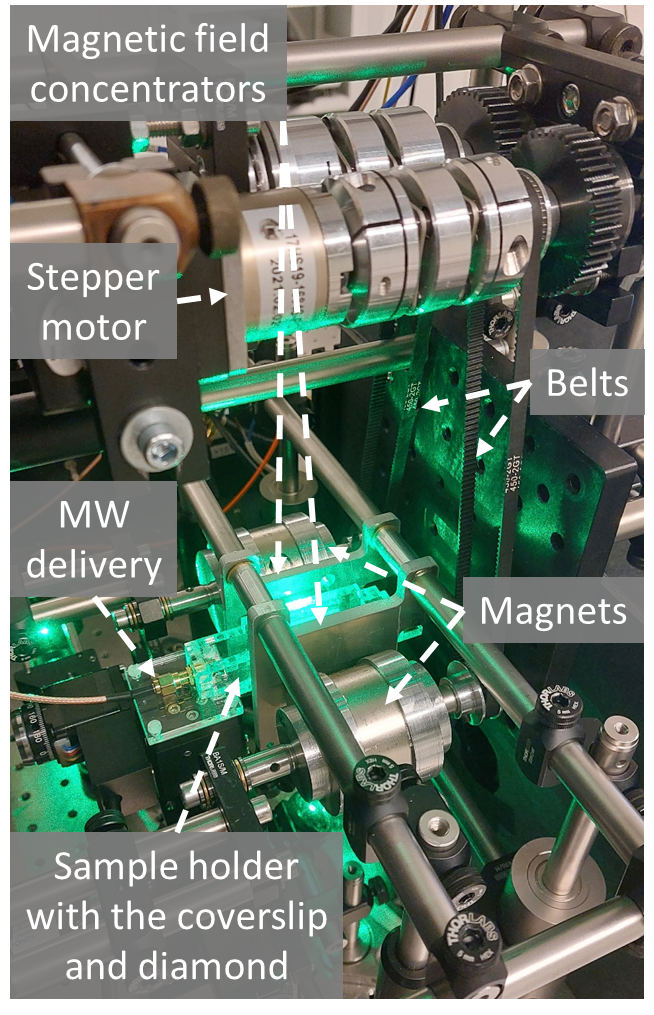}
    \caption{\textbf{Left}: Schematic of the experimental apparatus for widefield diamond microscopy, DAQ: data acquisition card; PC: personal computer. \textbf{Middle}: 3D model of the permanent magnet rotation mechanism. \textbf{Right}: The magnet setup in the laboratory. The ferrite concentrators are held in place by custom made aluminium holders, the coverslip with the diamond is placed between the magnetic field concentrators by a custom made sample holder.}
  \label{setup}
\end{figure*}

The experimental setup consists of two main parts:
the optical system to guide the exciting green laser light to the diamond sample and afterwards gather the red fluorescence from the NV centers in the diamond and guide it to a photodiode or a camera, and the electro-mechanical system for applying magnetic field by controling the rotation angle of the two cylindrical permanent magnets. The setup for widefield magnetic field imaging with the system for the application of a bias magnetic field is depicted in Figure~\ref{setup}.

The diamond we used as the magnetic field probe is an HPHT diamond with dimensions 2.0~$\times$~2.0~$\times$~0.06~mm and (110) surface polish. We performed \textbf{S}topping \textbf{R}ange of \textbf{I}ons in \textbf{M}atter or SRIM simulations~\cite{ziegler_srim_2010} to determine the implantation parameters required for the fabrication of a 200-nm-thick NV layer close to the diamond surface. The crystal was irradiated with $^{4}$He ions at three separate energies and doses: 6.0$\times 10^{12}$~He/cm$^2$ at 5~keV, 6.0$\times 10^{12}$~He/cm$^2$ at 15~keV, and 1.20$\times$ 6.0$\times 10^{13}$~He/cm$^2$ at 33~keV. After that the diamond was annealed for four hours at 800 C$^{\circ}$ and for four hours at 1100 C$^{\circ}$ under high vacuum to promote migration of vacancies to the substitutional nitrogen defects, as well as to allow the crystal to heal after irradiation.

The diamond is placed on a coverslip that is attached to a sample holder, that in turn is attached to a 3 axis translation stage enabling the sample movement and focus adjustment in an epifluorescent microscope. The NV excitation and fluorescence detection is performed through the same infinity-corrected 100$\times$ microscope objective with a numerical aperture of 1.25 (100x/1.25 Oil, $\infty$/0.17, ZEISS). The NV centers are exposed to 200~mW of green radiation guided by a multi-mode optical fiber and lens system from a Coherent Verdi V-18 laser. The NV fluorescence ($650\mbox{--}800$ nm) is separated from the green light by a dichroic mirror (Thorlabs DMLP567R) and imaged through a long-pass filter (Thorlabs FEL0600) on a sCMOS sensor of an Andor Neo 5.5 camera, or on a photodiode (Thorlabs PDA36A-EC). The field of view detected by the Andor camera detector region of 512$\times$512 pixels during measurements is an area of about $25\times25~\upmu {\rm m}^2$. The diffraction-limited spatial resolution defined by the NA of the microscope objective sets the fundamental limit of resolution $\sim 350$~nm, that is reduced by the air gap between the coverslip and the diamond to an estimate of $\sim 700$~nm.

The MW field, necessary for the ODMR measurements, is produced by a SRS SG386 generator. The MW frequency is slowly swept using the analog voltage of a sawtooth waveform, which is delivered from a data acquisition card (NI USB-6001). The amplified MW field (Mini Circuits ZVE-3W-83+ or ZVE-3W-183+ amplifier) is delivered to the diamond by a copper wire with a diameter of 50~$\upmu$m. To reach some of the necessary frequency regions we also used a frequency doubler Mini Circuits ZX90-2-24-S+. The MW frequency is swept across a central frequency of an ODMR ($\vert0\rangle\leftrightarrow\vert+1\rangle$ or $\vert-1\rangle$  transition) profile with a deviation of $\pm$15 MHz. The MW sweep is triggered by a pulse from the sCMOS camera, which records 40 frames per sweep. The acquisition time of one frame ($512\times512$ pixels) is 7 to 9 milliseconds (depending on the set exposure time). Five series of frames are averaged in the camera memory before being read out by a LabVIEW interface. Typically, for magnetic field gradient measurements we averaged about 3000 measurements. As a result, we obtained NV fluorescence images that reveal an ODMR shape for each pixel. In post-processing, we fit the obtained ODMR profiles pixel-wise with a Lorentzian function to obtain a two-dimensional map of the ODMR central frequencies for both $\vert0\rangle\leftrightarrow\vert+1\rangle$ and $\vert-1\rangle$ transitions separately. After that, the obtained values for each transition for each corresponding pixel is subtracted from each other, the results can then be interpreted as magnetic field values. In this way a magnetic field image is constructed.

The magnets used for this setup are custom made diametrically magnetized (the magnetization vector is perpendicular to the principal cylindrical axis of the magnets) N50 magnet cylinders (from NINGBO ZHAOBAO MAGNET CO., LTD.), 30~mm in height and 50~mm in diameter. There is a 44~mm gap between the outer edges of the permanent magnets. The magnetic field concentrators 30~mm $\times$ 50~mm $\times$ 5~mm made of ferrite MN60 (The National Magnetics Group. Inc.) are placed at a distance of 4~mm from the outer edges of the permanent magnets (with their longer edge in the direction of the permanent magnet cylindrical axis) leaving a free space of 26~mm between the ferrite concentrators, where the diamond sensor is placed at the center point between the permanent magnets, where the magnetic field produced by this system is the most homogeneous, see FIG. \ref{setup}. The magnets are fixed in space so that the distance between them can not change. The magnet mounts allow them to spin around their respective main cylindrical axes.

To diminish the magnetic field gradient that the diamond sensor could be exposed to, the magnet system is placed around the diamond in a way that the principal cylindrical axis of the magnets are parallel to the plane of the coverslip, see FIG. \ref{setup}. In this way only the 200~nm thin NV layer in the diamond is sensing the the magnetic field gradient that is more pronounced in the direction perpendicular to the diamond surface (see FIG.~\ref{FEMM}). And, as the magnetic field lines have a relatively large curvature radius, it can be considered that within the field-of-view and the depth of NV layer (200~nm) the magnetic field is relatively homogeneous. It has to be noted that a situation where the principal cylindrical axis of the magnets are perpendicular to the plane of the coverslip might seem to be very similar, but in practice it is not so, the reason is clearly visible in FIG.~\ref{FEMM} b) where the gradients along the \textbf{Y} axis is much less pronounced than along the \textbf{X} axis.

The permanent magnet system is aligned and controlled in the following way: the rotation (the angular position) of the magnets is controlled by a stepper motor with a reduction unit (resulting in 54920 steps per revolution) controlled by a custom software. The stepper motor drives a shaft which is connected to another separate shaft via two equal spur gears, ensuring that the two shafts rotate in the opposite directions (FIG.~\ref{setup} middle). This is necessary to enable the system to reach the two extreme points: 1) to reach the maximum magnetic field, magnet poles are aligned with the \textbf{Y} axis and face in the same direction, 2) to reach the minimum value of the magnetic field, magnet poles are aligned with the \textbf{X} axis and face in the opposite directions. Such a system also makes it possible to achieve all intermediate values of the magnetic field between the two extreme cases. Further this stepper motor, shaft and gear aggregate is connected to the constructions holding the magnet cylinders themselves via a timing pulley and timing belt system. The magnetic field produced in between the two magnets can then be easily aligned with one of the NV axes in the diamond which is surface polished along the (110) direction.

Such a system have several benefits: counter-rotating gears and magnets ensure that the stepper motor has to apply only a small amount of force as the magnetic force created by the two magnets in perspective of the magnet rotation always works against each other; changes in the magnetic field requires low current and voltage; timing belts allow to position the stepper motor relatively far from the center of the experimental system (in our case $\approx$~20 cm away from the magnets); the stepper motor with the reduction unit allows for very precise control of the angle of the magnets.

The diamond sample is placed in the free space gap in between the ferrite concentrators at the center point between the two permanent magnets. The diamond is held in place on a coverslip by a copper wire used for the MW delivery. The wire is tensioned in order for it to lightly press the diamond against the coverslip to hold it in place (still allowing for some rotation of the diamond for alignment purposes) as well as to ensure the minimal possible distance between the wire and the NV layer. The magnetic field images obtained using this experimental setup are used to determine the uniformity of the magnetic field between the two permanent magnets.

\section{Results}
\begin{figure*}[h]
  \begin{center}
    \includegraphics[trim=0 0 0 0,clip,width=0.99\textwidth]{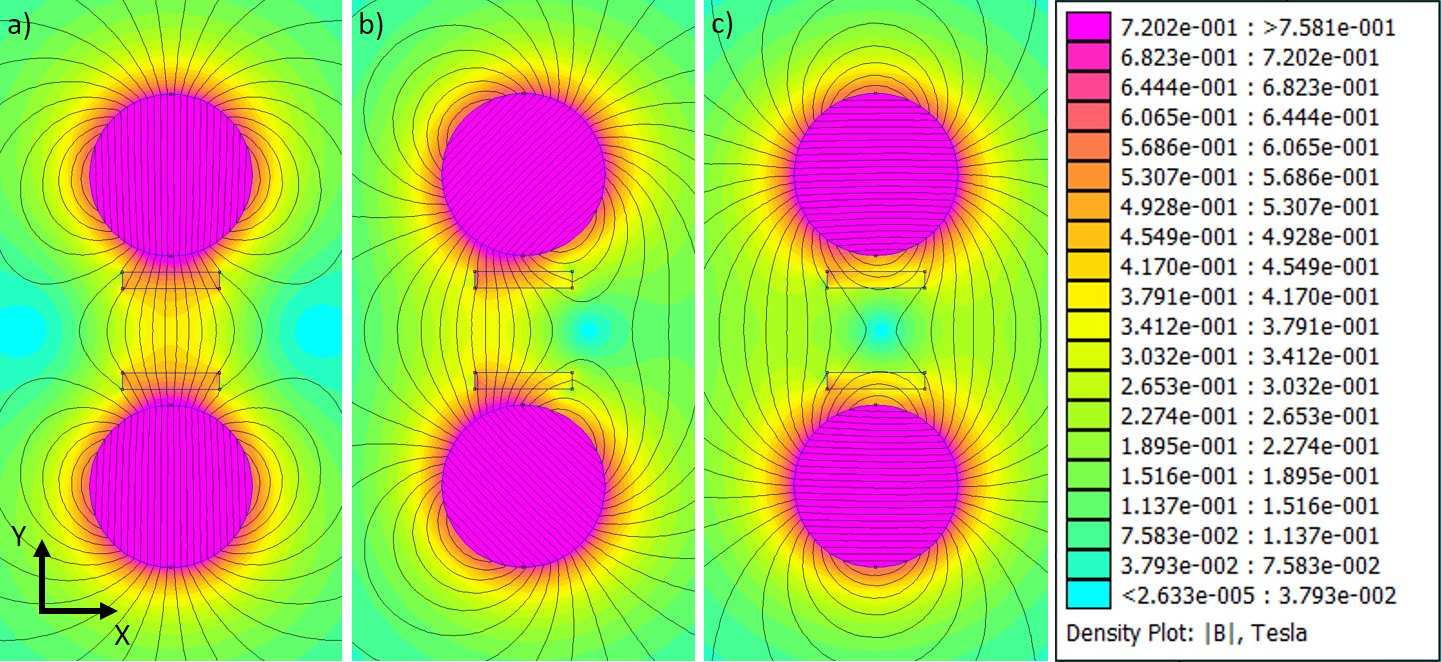}
  \end{center}
  \caption{A 2D simulation of the magnetic field distribution performed using a finite element simulation software (FEMM) showing the two permanent magnets with the ferrite concentrators between them. The observation for this simulation is done in the direction of the magnets cylindrical axis, this allows to see a magnetic field gradient in the \textbf{X} direction between the magnets for multiple magnetic field configurations. Due to this gradient, the coverslip with the diamond is placed perpendicular to the \textbf{X} direction. The configurations of the two permanent magnets used in the simulations are: a)~$\uparrow$~$\uparrow$, b)~$\nearrow$~$\searrow$, c)~$\rightleftarrows$.}
  \label{FEMM}
\end{figure*}

\begin{figure*}
  \begin{center}
    \includegraphics[width=0.98\textwidth]{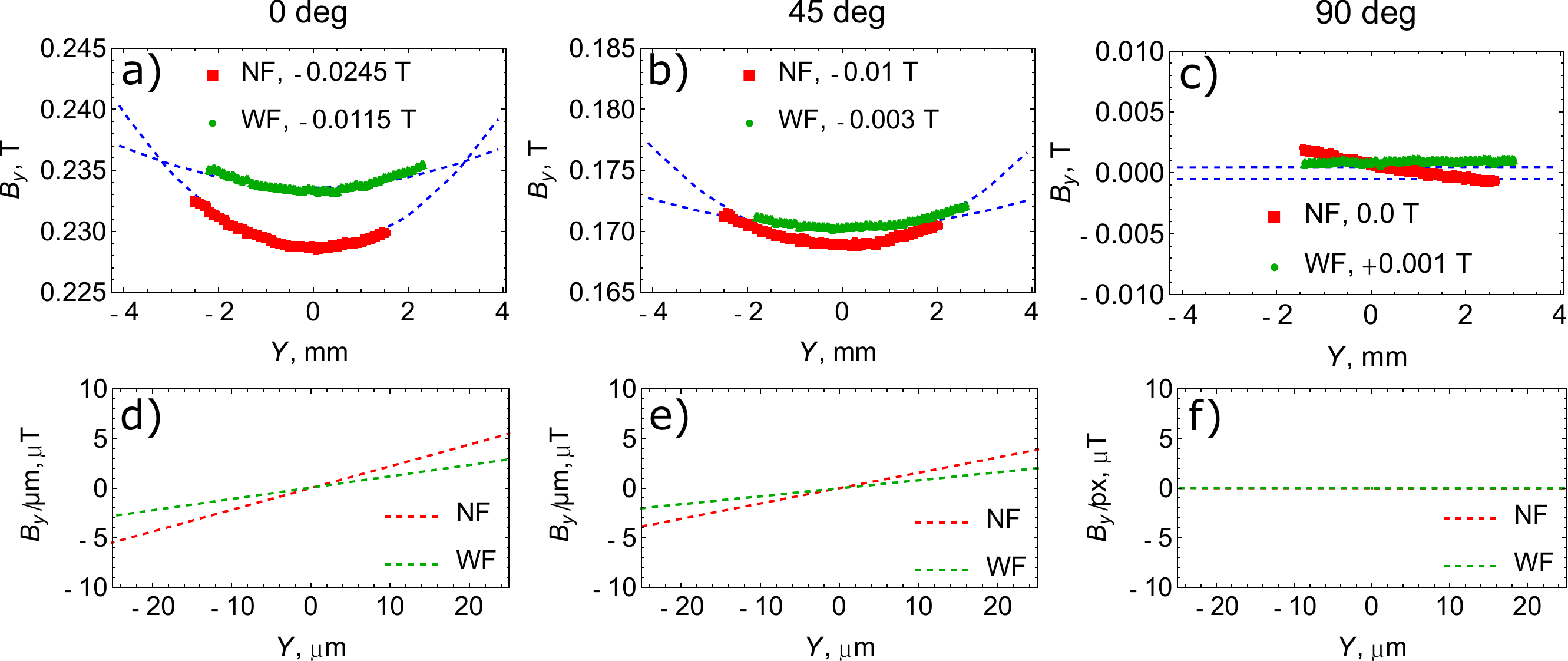}
  \end{center}
  \caption{Simulations and measurements done by a Hall probe. These line plots show the magnetic field distribution along the \textbf{Y} direction between the two magnets for three different configurations of the two permanent magnets: a)~$\uparrow$~$\uparrow$, b)~$\nearrow$~$\searrow$, c)~$\rightleftarrows$. Case with field concentrators denoted as \textbf{WF} or With Ferrites versus case without field concentrators denoted as \textbf{NF} or No Ferrites.}
  \label{Comsol vs Hall}
\end{figure*}

To characterize the homogeneity of the magnetic field at the center between the two permanent magnets with the magnetic field concentrators, we first did magnetic field simulations by using the \textbf{F}inite \textbf{E}lement \textbf{M}ethod \textbf{M}agnetics or FEMM simulation software (FIG. \ref{FEMM}) and COMSOL simulations (FIG. \ref{Comsol vs Hall}) for three separate cases:

\begin{enumerate}[label=\textbf{\Alph*)}]
    \item both the magnetization directions are facing in the same direction ($\uparrow$~$\uparrow$), producing the maximum possible magnetic field,
    \item the magnetization directions both make a 45$^{\circ}$ angle with respect to the \textbf{Y} direction ($\nearrow$~$\searrow$), producing an intermediate magnetic field value,
    \item the magnetization directions are opposite to one another ($\rightleftarrows$), producing a minimum magnetic field.
\end{enumerate}

Numerical simulations from FEMM (FIG.~\ref{FEMM}) show that at the center between the magnets a gradient along the \textbf{X} axis exists in all cases except when the magnets produce a close to zero magnetic field or are aligned for the maximum magnetic field. However, gradients along the \textbf{Y} axis are much lower. From the FEMM simulations we obtained the following magnetic field values in the centre of the device: for case \textbf{A)} 406~mT, for case \textbf{B)} 296~mT, and for case \textbf{C)} 1.0~mT.
To clearly demonstrate the difference between the magnet systems with and without field concentrators and to analyze the magnetic field in the centre of the magnet system along the \textbf{Y} axis (axis that is expected to give the largest gradient value in YZ plane) we made measurements using a Hall probe on a positioning platform (allowing a 4~mm travel range) and made COMSOL simulations for cases \textbf{A)}, \textbf{B)} and \textbf{C)} (FIG.~\ref{Comsol vs Hall}).

Firstly, experimentally measured results, using a Hall magnetic field probe, compared to the magnetic field distribution simulations (FIG. \ref{Comsol vs Hall} a)-b)), show good agreement between the experiment and the predicted behaviour of the magnet system from the simulations. In this case the maximum deviation between the measurement and simulation reached 24.5~mT, in comparison of the FEMM simulations, where the simulations gave roughly two times larger magnetic field amplitudes than the measured ones (even after diligent debugging of the model input data). We explain this with the difference in the simulation mechanism, namely, in COMSOL we simulate a full 3D problem, and then look at a single slice of the solution, while in the case of FEMM the simulations are done just for a single slice which, in principle, gives rise to a number of assumptions which are not always justified. However, we have no reason to believe that the FEMM simulations are incorrect in the perspective of the representation of the magnetic field distribution.

Secondly, measurements and simulations show the benefit of having the field concentrators in all cases (FIG.~\ref{Comsol vs Hall} a)-b), simulations and measurements denoted as WF or With Ferrites versus simulations and measurements denoted as NF or No Ferrites). The field concentrators straighten out the magnetic field lines with respect to the coverslip/diamond plane, lessening the magnetic field gradient in the \textbf{Y} direction. Besides, concentrators compensate for small deviations from the perfect geometry of the magnet position (pitch, roll or imperfections of angular rotation of one magnet with respect to the other), that can easily occur in real-life systems.

Thirdly, FIG.~\ref{Comsol vs Hall} d)-f) present the derivative of the data obtained by the COMSOL simulations, giving the magnetic field gradient per $\upmu$m for each magnet orientation. Again, the benefit of having the field concentrators (results denoted as WF) is clearly visible, yielding a gradient with roughly twice as less amplitude compared to the case without concentrators (results denoted as NF).

\begin{figure}[h]
  \begin{center}
    \includegraphics[width=0.49\textwidth]{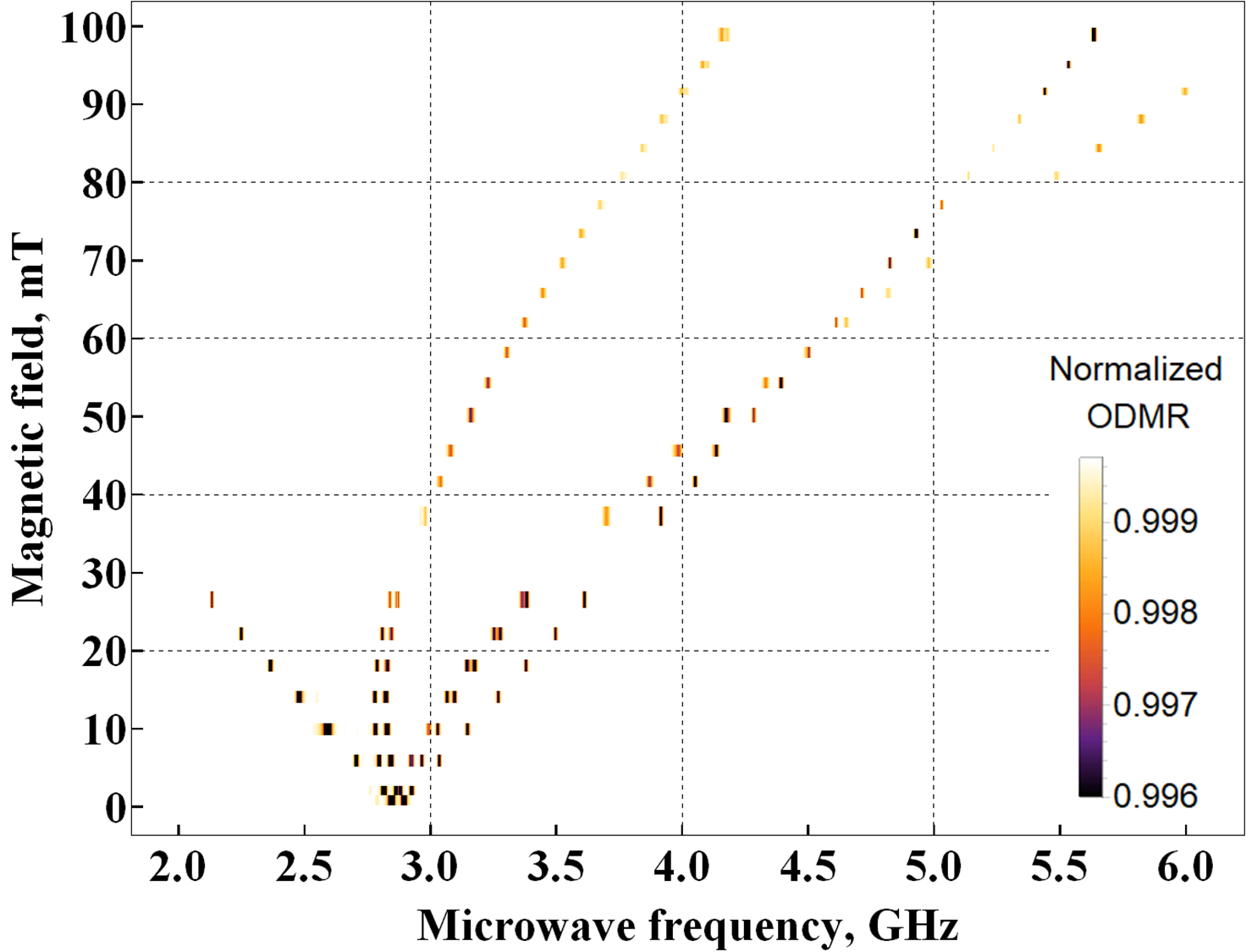}
  \end{center}
  \caption{ODMR from the NV centres giving the average ODMR frequency value over the illumination region of $\approx$ 30~$\upmu$m~$\times$~30~$\upmu$m$^2$. The permanent magnet rotation angles were changed from 0$^{\circ}$ to 24$^{\circ}$ resulting in a magnetic field change from close to zero to 100~mT. The ODMR measurements were done keeping the three ODMR signal components that are not aligned with the magnetic field in the MW scan range to see how they change depending on the magnet angle, how good they are overlapping.}
  \label{odmr-vs-angle}
\end{figure}

To test the experimental set-up on the micro scale we performed ODMR measurements in two measurement modes: measuring the ODMR signal with a photodiode for magnetic fields up to 100~mT, and magnetic field imaging measurements for three magnetization directions \textbf{A)}, \textbf{B)} denoted in the start of this chapter and a configuration producing a magnetic field of 28~mT. The results from the measurements with a photodiode are depicted in FIG.~\ref{odmr-vs-angle}, giving the average ODMR value over the illumination region of $\approx$ 30~$\upmu$m~$\times$~30~$\upmu$m for a range of magnetic field values.

One of the things that can be seen in FIG.~\ref{odmr-vs-angle} is that at the minimum magnetic field value the ODMR profile is still split, meaning that the magnetic field is not zero, it is estimated to be approximately 0.95~mT. Another feature that can be seen from FIG.~\ref{odmr-vs-angle} is that at low magnetic fields (up to 40~mT) the bias magnetic field is not perfectly aligned with one of the NV axis, as the resonances created by the NV centers that are not aligned with the magnetic field are split in two. This occurs simply due to the fact that the fine tuning of the diamond orientation was done at $\approx$~70~mT. A coarse alignment was done by rotating the diamond on the surface of the coverslip and fine adjustments were done by slightly changing the angle of the magnetic field concentrators in order to compensate for any magnetic field in the vertical direction (\textbf{X} axis) with respect to the diamond. A small misalignment of the diamond most probably occurs and also laboratory fields, that change when the magnets are rotated (mainly because of the microscope objective and its holder, that are not completely nonmagnetic), change the direction of the bias magnetic field.

\begin{figure*}[h]
  \begin{center}
    \includegraphics[width=0.98\textwidth]{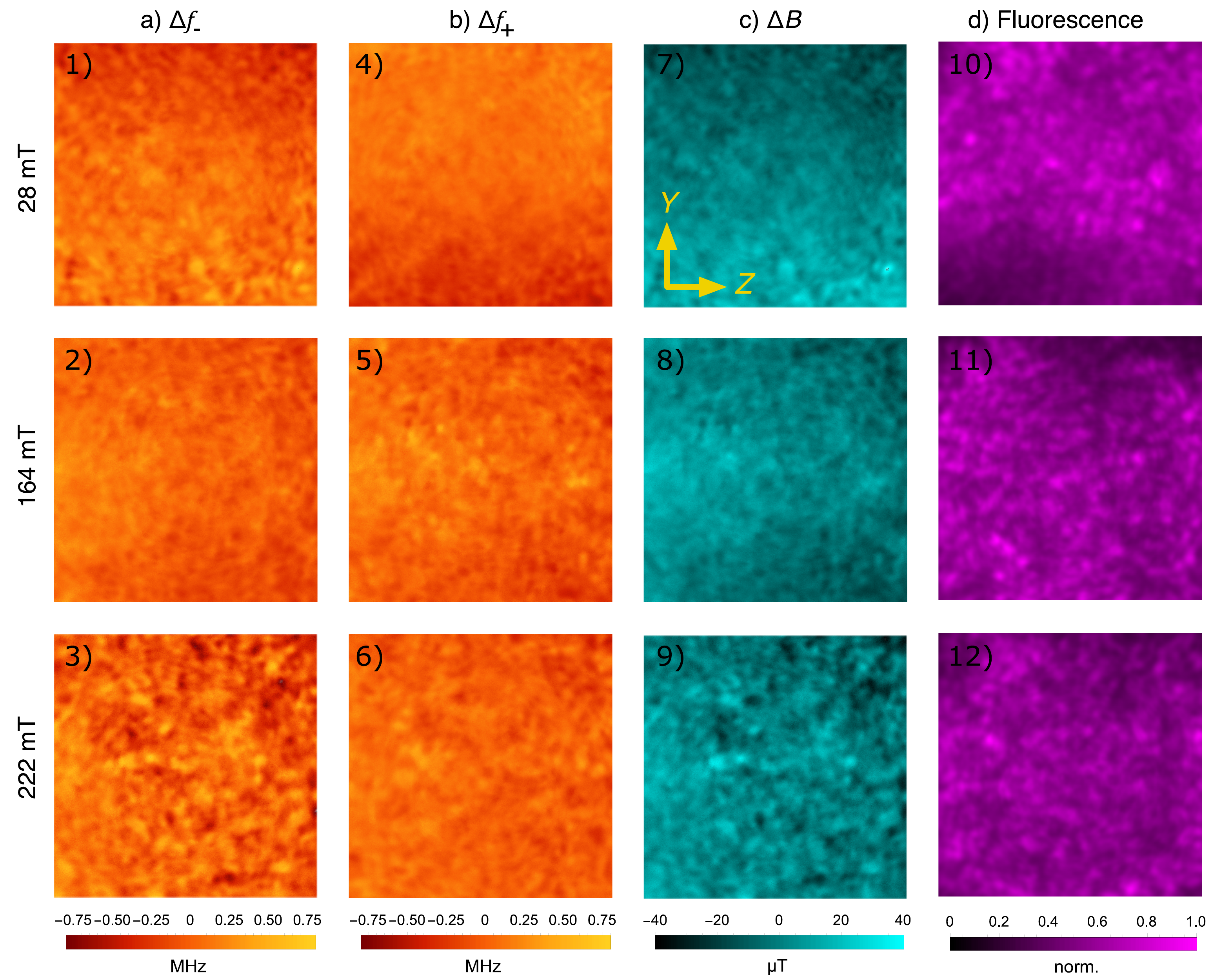}
  \end{center}
  \caption{\textbf{Top}: Magnetic field images at a bias magnetic field of 28~mT. \textbf{Middle}: Magnetic field images at a bias magnetic field of 164~mT. \textbf{Bottom}: Magnetic field images at a bias magnetic field of 222~mT. From left to right: ODMR frequency shift map obtained from the $\vert 0\rangle\longrightarrow\vert -1\rangle$ transition, ODMR frequency shift map obtained from the $\vert 0\rangle\longrightarrow\vert +1\rangle$ transition, magnetic field map obtained using the difference of the frequency shifts from both transitions ($\vert 0\rangle \longrightarrow\pm 1\rangle$) and dividing the difference with the coefficient of 28.03~MHz/mT, the normalized fluorescence map from both of the transitions.}
  \label{28mT}
\end{figure*}

To determine the spatial distribution of the magnetic field gradient on a micro scale we performed magnetic field imaging measurements for the three magnetization directions \textbf{A)} producing a 222~mT field, \textbf{B)} producing a 164~mT field denoted in the start of this chapter and a configuration producing a 28~mT field as zero and close to zero magnetic fields can not be measured by the method we are using. It has to be noted that the maximum magnetic field value is a bit different than in measurements with the Hall probe because the distance between the concentrators was slightly changed. The obtained results can be seen in Figure~\ref{28mT}. In all measurements the results are depicted in the following way: the data in each row represents one measurement set at the magnetic field value denoted at the left side of the row; columns a) and b) represent the ODMR maps obtained for separately measured transitions $\vert 0\rangle\leftrightarrow\vert +1\rangle$ and $\vert-1\rangle$; column c) presents the obtained magnetic field maps; d) presents the total fluorescence distribution normalized to the maximum value of the fluorescence over the field of view.

The ODMR maps in columns a) and b) in Figure~\ref{28mT} present the resonance frequency maps, where the ODMR frequencies are given a color coding for creation of 2D maps. In the case of FIG.~\ref{28mT} measurements at 28~mT, the ODMR gradients in 1) and 4) present a shift in opposite directions, but in measurements at 164~mT 2) and 5) and at 222~mT 3) and 6) the frequency shifts have the same overall shape. This is expected and can be easily explained with the bias magnetic field in perspective of the NV centre energy scheme: in the case of FIG.~\ref{28mT} 1) and 4) the measurements are done at frequencies
that are well below the GSLAC~\cite{auzinsh_hyperfine_2019}, meaning that by increasing the magnetic field the transition $\vert 0\rangle\longrightarrow\vert -1\rangle$ moves towards lower frequencies, but $\vert 0\rangle\longrightarrow\vert +1\rangle$ moves towards higher frequencies. However, in the cases of FIG.~\ref{28mT} 2) and 5), and FIG.~\ref{28mT} 3) and 6), the central ODMR frequencies are above the GSLAC, and thus by increasing or decreasing the magnetic field, both transitions experience frequency shifts in the same positive or negative direction.

\begin{figure}[h]
  \begin{center}
    \includegraphics[width=0.48\textwidth]{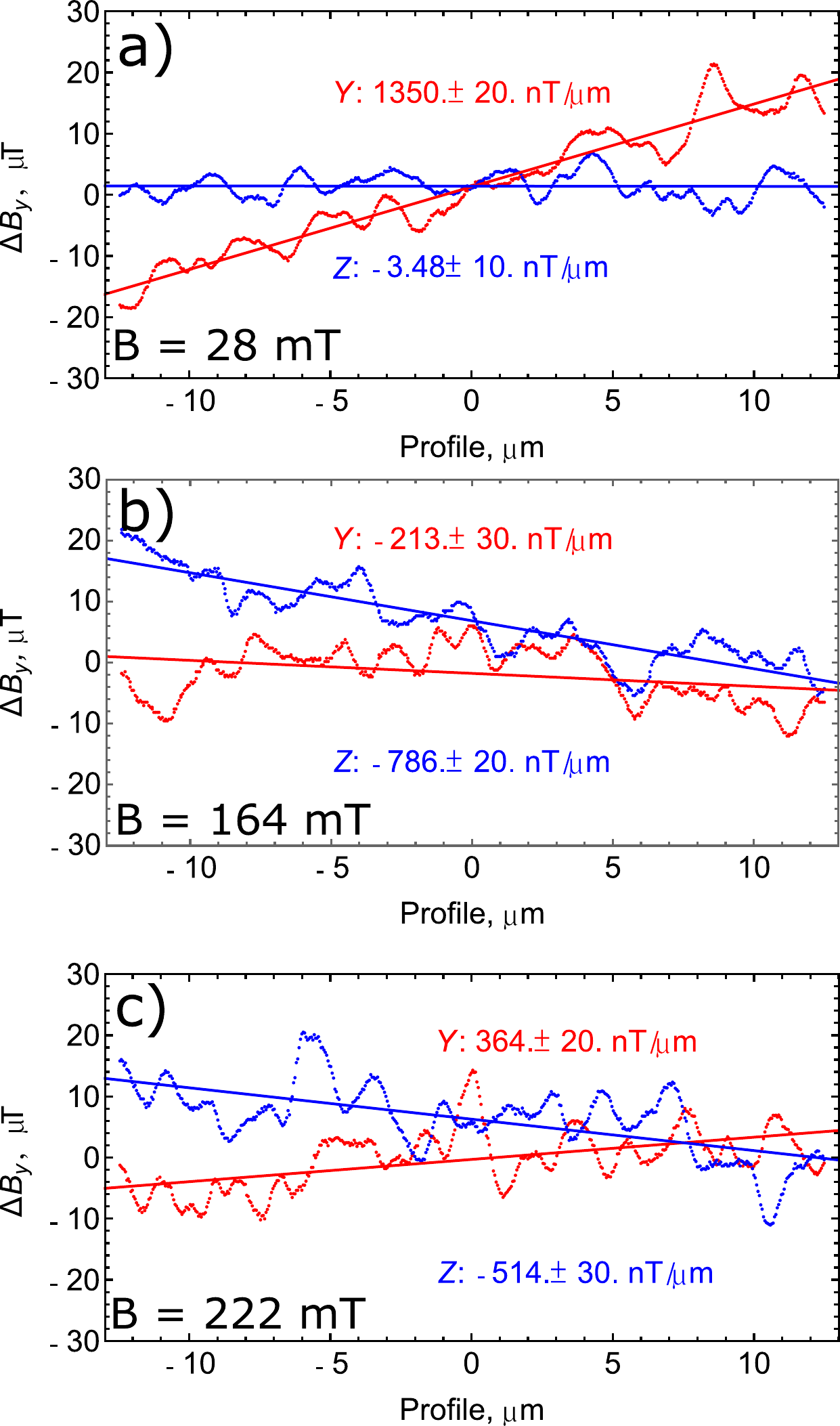}
  \end{center}
  \caption{Experimental results of the magnetic field gradient (with ferrites between the permanent magnets). These line plots representing the magnetic field gradients are constructed in the following way, both profiles Y and Z are constructed by taking data from the middle part of the magnetic field profiles presented in column c) of FIG.~\ref{28mT}. To lessen the impact of the noise we averaged 30 adjacent pixel rows for Z profile, and 30 adjacent columns for Y profile. The straight lines are linear fits of the data, the slope values give the change in magnetic field per $\upmu$m for each direction.}
  \label{results-profiles}
\end{figure}

The magnetic field maps (column c) in FIG.~\ref{28mT}) are obtained in the following way, first, the two corresponding ODMR maps are subtracted from each other (or added in the cases where the magnetic field is above 102.4~mT). This is done to get rid of signals that could potentially occur due to strain and temperature related effects, as these effects shift both resonance frequencies in the same direction, while magnetic field related effects shift the resonances in the opposite directions. After the subtraction, the obtained frequencies are divided by 28.025, as the transition frequency for the NV centres aligned with the magnetic field is linear and changes by $28.025$~MHz/mT. After that the 2D magnetic field map is given color coding. For in depth analysis we also constructed line plots representing the magnetic field gradients, see FIG.~\ref{results-profiles}. The results obtained give a convenient mode of assessment of the field gradient over the field of view, as well as gradient determination over one camera pixel. These line plots representing the magnetic field gradients are constructed in the following way, both profiles $Y$ and $Z$ are constructed by taking data from the middle part of the magnetic field profiles presented in column c) of FIG.~\ref{28mT}. To lessen the impact of the noise we averaged 30 adjacent pixel rows for $Z$ profile, and 30 adjacent columns for $Y$ profile. The straight lines are linear fits of the data, the slope values give the change in magnetic field per $\upmu$m for each direction.

One would expect that larger bias magnetic fields create larger magnetic field gradients (in the means of amplitude), however in our case the magnetic field gradient in the \textbf{Y} direction is largest at 28~mT and smallest at 164~mT. The magnetic field gradient in the \textbf{Z} direction is very low at 28~mT and is the highest at 164~mT. There are two main reasons for varying magnetic field gradient amplitudes, first, the way how we fine-tune the magnetic field direction relative to the diamond is by making small adjustments to the concentrator pair angle, thus this potentially can move the magnetic centre of the set-up. In other words, looking at FIG.~\ref{Comsol vs Hall} non-zero field cases, the concentrator pair adjustments move the measurement spot to the left or right of the centre (0~mm) position, meaning that the measurements are not done at the bottom of the U-shape, but on one of the sides, resulting in a larger gradient. Second, as the microscope objective and its holder, are not completely nonmagnetic, and, by rotating the magnets, these parts experience and generate some magnetic field of their own, thus slightly changing the direction of the total field felt by the diamond. And it is expected, that these small deviations have a larger effect at smaller bias fields when small perturbations have a more noticeable effect on the general direction of the magnetic field.

\begin{figure}[ht]
  \begin{center}
    \includegraphics[width=0.4\textwidth]{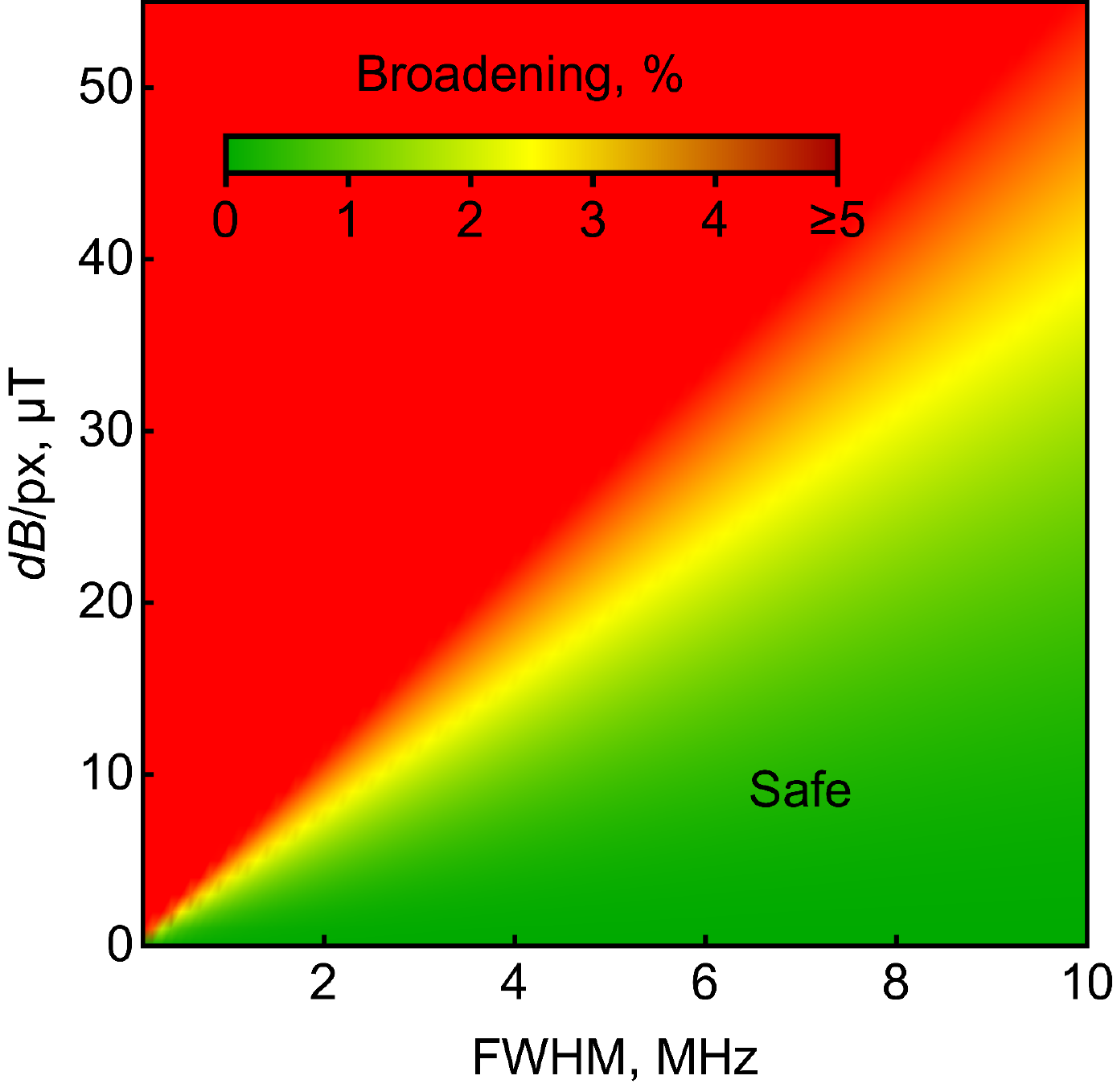}
  \end{center}
  \caption{The broadening of the FWHM of an ODMR profile in relation to the magnetic field gradient over one detection pixel ($0.5\times0.5$~$\upmu$m$^2$). The color scale indicates the regions where the broadening is negligible (green) and regions where broadening will have a noticeable effect on the measurement (red).}
  \label{gradients}
\end{figure}

To give another perspective on the obtained results we estimated the impact of a magnetic field gradient on the FWHM of the ODMR profile within one detection pixel of the magnetic microscope. Here we assumed that the average size of the detection pixel is $0.5\times0.5$~$\upmu$m$^2$. Based on this one can estimate the maximal magnetic field gradients acceptable for optimal magnetic field imaging measurements, see FIG.~\ref{gradients}. By looking at this color map we can see that even the largest gradient measured with this system $\approx$~1.3~$\upmu$T/$\upmu$m or $\approx$~0.65~$\upmu$T/px at an average FWHM of 6~MHz of an ODMR profile is well within the zone, where the ODMR signal broadening is below 1\%.

\begin{figure}
      \includegraphics[width=0.4\textwidth]{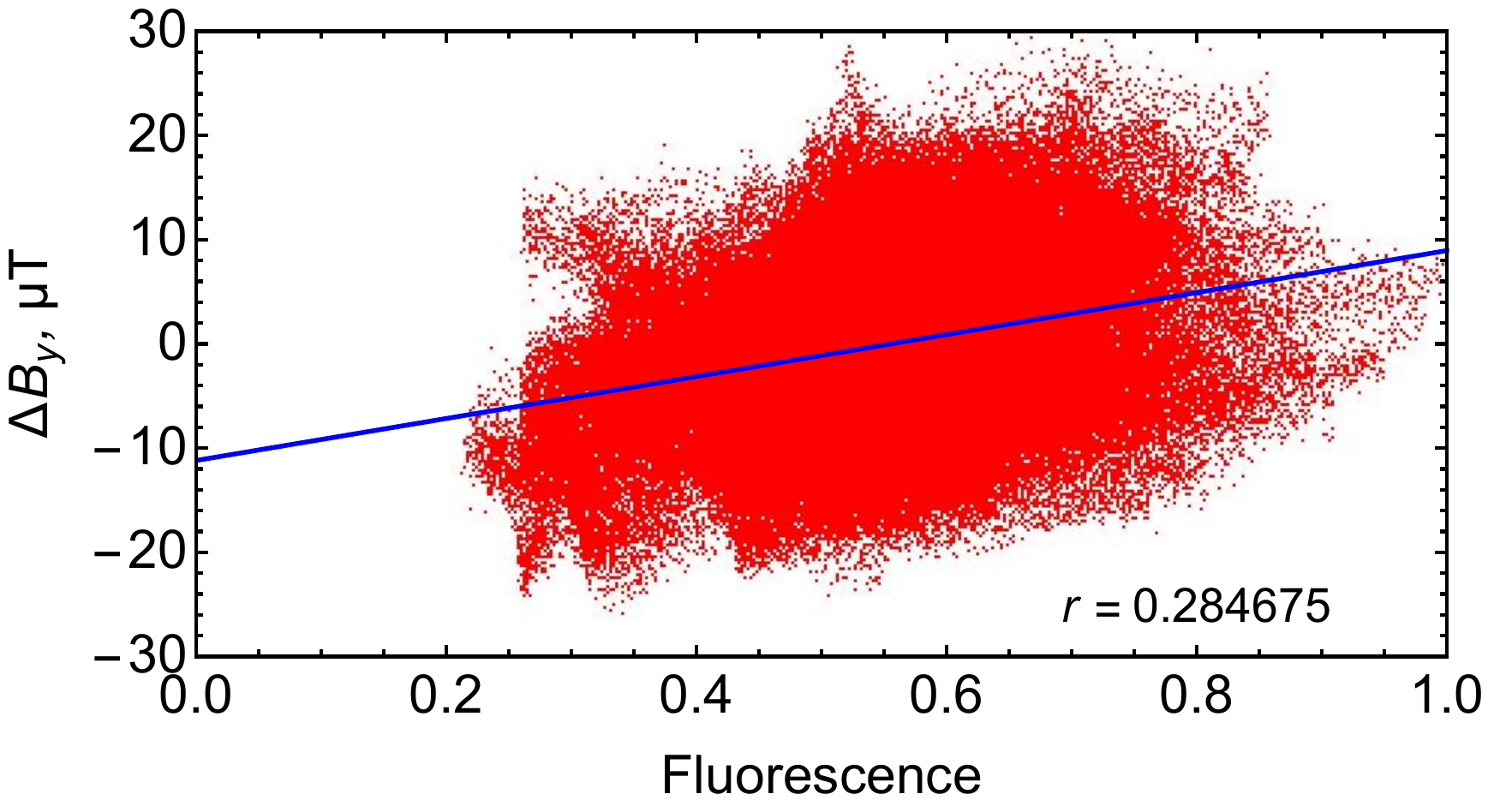}
     \caption{Magnetic field $\Delta B_y$ (Fig.~\ref{28mT} 8)) vs. normalized fluorescence (Fig.~\ref{28mT} 11)) plotted pixel-wise alongside with a linear fit and Pearson correlation in the inset.} 
      \label{correlation}
\end{figure}

As the magnetic field maps contain some noise structure, we tried to find their origin. The plausible causes in our understanding could be: NV layer illumination related artefacts in combination with ODMR profile fitting artefacts, and NV distribution. To analyse the possible NV layer illumination contribution to the noise structure we created a correlation graph for the 164~mT measurement where the magnetic field map (FIG.~\ref{28mT} 8)) is correlated with the fluorescence map (FIG.~\ref{28mT} 11)), see FIG.~\ref{correlation}. As can be seen from this graph, there is some correlation between the fluorescence distribution and the magnetic field distribution, meaning that the fluorescence distribution to some extent impacts the noise in the magnetic field images. This might be connected with ODMR profile fitting related artefacts. As the shape of the ODMR profile at 3000 averages is still a little rough, and part of the noise signal might come from small spatial illumination fluctuations, it is possible that a fit with a single Lorentzian profile could give small deviations in the resonance frequency value. Moreover, if the noise structure is caused by this effect, it is easy to explain the sharp transitions between neighbouring regions (pixels) that should have relatively smooth transitions in the case of optical signals coming from the NV fluorescence region (due to diffraction limited spatial resolution).
Artefacts due to the NV distribution are not very likely, as in our case the spatial resolution of the system is around 0.7~$\upmu$m, meaning that if one can see spatial fluctuation in the NV distribution, these fluctuations should be relatively sharp. But taking into account the He ion implantation doses and annealing process it is unlikely.

The central frequencies of the ODMR measurements, corresponding frequency ranges and magnetic field values, and gradient values vary by the applied bias field. The values of these parameters are compiled in TABLE~\ref{results-table}.

\begin{table}[h!]
\centering
\begin{tabular}{|p{0.1\linewidth}|p{0.1\linewidth}|p{0.1\linewidth}|p{0.1\linewidth}|p{0.07\linewidth}|p{0.07\linewidth}|p{0.15\linewidth}|p{0.15\linewidth}|}
\hline
$f_{-1}$, MHz & $\Delta f_{-1}$, MHz & $f_{+1}$, MHz & $\Delta f_{+1}$, MHz & $B$, mT & $\Delta B$, $\upmu$T & $\Delta B_Y$, ppm/$\upmu$m & $\Delta B_Z$, ppm/$\upmu$m \\ [0.5ex]
\toprule
2080 & 1.45 & 3659 & 0.85 & 28 & 70 & 48 & 0.12 \\\hline
1740 & 0.85 & 7479 & 1.05 & 164 & 56 & 1.3 & 4.8 \\\hline
3354 & 3.6 & 9091 & 1.7 & 222 & 170 & 1.6 & 2.3 \\\hline
\end{tabular}
\caption{Results: ODMR frequency variations ($\Delta f_{\pm1}$) for both $\vert 0\rangle\longrightarrow\vert +1\rangle$ and $\vert 0\rangle\longrightarrow\vert -1\rangle$ transitions. Using the results from both of these transitions the variation of the magnetic field over the field of view ($\Delta B$) is also obtained. The gradient values $\Delta B_Y$ and $\Delta B_Z$ were estimated by dividing the linearly fitted values in FIG. \ref{results-profiles} with the corresponding bias magnetic field values.}
\label{results-table}
\end{table}








\section{Conclusions}
In this study we have given a blueprint for a setup for a compact, tunable and uniform magnetic field source, using cylindrical diametrically magnetized permanent magnets with field concentrators that can be used for various microscopy applications. This electromechanicaly controllable system not only gives a compact, inexpensive and energy efficient alternative for electromagnets, that is not prone to heating up during usage, and does not rely on the quality of power supplies, but also allows modifications to fit different needs. Moreover, the design presented here at created 222~mT magnetic field gives a magnetic field gradient of 2~ppm/$\upmu$m or 0.5~$\upmu$T/$\upmu$m. In comparison to a magnetic field setup using just one disc magnet, commonly used in similar experiments to generate the bias magnetic field, where the typical obtained magnetic field gradients is about 2-3~$\upmu$T/$\upmu$m at magnetic fields of 20-30~mT~\cite{berzins_characterization_2021}.

Rotation of both of the permanent magnets around their main cylindrical axes allows for rapid changes from a close to zero magnetic field to a relatively high magnetic field, giving a useful tool for hysteresis measurements, usually requiring a combination of relatively large magnetic fields with the possibility to gradually increase or decrease the magnetic field, as well as the ability to change the direction of the magnetic field. Furthermore, this same design potentially allows for measurements requiring a dynamically changing magnetic field (for instance, magnetic relaxation measurements), yet rapid full revolution rotations of the magnet cylinders would cause changing mechanical stresses experienced by the whole setup, that would need to be accounted for.

The maximum magnetic field produced by the system presented in this work (222.1~mT) is mainly limited by the distance between the magnets, the field concentrators and the centre of the device, and by the magnet properties. The main obstacle in similar setups is the diameter of the microscope objective (in our case 25~mm). One immediate solution could be the usage of long working distance objectives~\cite{berzins_surface_2021}, that, on one hand, allow to put the microscope objective outside the volume between the concentrators, but, on the other hand, dramatically reduce the spatial resolution and the light collection efficiency of the system.

The minimum magnetic field that the current experimental setup device was able to produce was $\approx$~0.95~mT. This value agrees well with the 2D magnetic field simulations (FIG. \ref{FEMM} c)). A conclusion that can be drawn from measurements and simulations is that this type of system can not reliably generate a zero magnetic field unaided, due to stray magnetic fields (the Earth or laboratory) that can easily change the field properties, and even in the best case scenario the zero field region is very narrow. This is not a problem for NV ensemble based measurements, as low magnetic fields cause a degeneration of the energetic levels of the differently oriented NV centres~\cite{mittiga_imaging_2018}, thus making the magnetic field measurements based on ODMR position determination challenging.

The presented magnetic field source design is not limited to NV based measurements, as very similar requirements for a magnetic field source can be found in other methods, like Magneto Optic Kerr Effect (MOKE) measurements~\cite{kim_extreme_2020}.

\section{Acknowledgements}
A. Berzins, H. Grube and R. Lazda acknowledges support from Latvian Council of Science project lzp-2020/2-0243 "Robust and fast quantum magnetic microscope with concentrated bias field". I. Fescenko acknowledges support from ERAF project 1.1.1.5/20/A/001. We also acknowledge LLC "MikroTik" donation projects, administered by the UoL foundation: "Improvement of Magnetic field imaging system" for opportunity to significantly improve experimental setup, "Simulations for stimulation of science" for opportunity to acquire COMSOL licence, and project "Annealing furnace for the development of new nanometer-sized sensors and devices".

\bibliography{references_zotero.bib}
\bibliographystyle{apsrev}

\end{document}